\documentclass{article}

\usepackage{graphicx}
\pdfoutput=1
\usepackage{amsmath,amssymb}
\usepackage{geometry}                
\usepackage[parfill]{parskip}    
\usepackage{amssymb}
\usepackage{epstopdf}
\usepackage{amsbsy}
\usepackage{amsmath}
\usepackage{tabularx}
\usepackage{longtable}
\usepackage{subfigure}
\usepackage{tikz}                                                                                                                                                                                                               
\usepackage{appendix}
\usepackage{natbib}
\usepackage{setspace}
\usepackage{authblk}
\usetikzlibrary{patterns}
\usetikzlibrary{calc}
\DeclareGraphicsExtensions{.pdf,.png,.jpg}

\title{Subglacial hydrology as a control on emergence, scale, and spacing of ice streams.}

\author[1,2]{T. M. Kyrke-Smith}
\author[2]{R. F. Katz}
\author[3,4]{A. C. Fowler}
\affil[1]{British Antarctic Survey, Cambridge, U.K.}
\affil[2]{Department of Earth Sciences, University of Oxford, Oxford, U.K. }
\affil[3]{MACSI, University of Limerick, Limerick, Ireland.}
\affil[4]{OCIAM, University of Oxford, Oxford, U.K.}

\date{}

\begin{document}
\maketitle

%
%


\begin{abstract}
Observations have long associated ice streams with the presence of meltwater at the bed. More recently, theoretical models have been able to reproduce ice-stream behaviour as a consequence of the coupled dynamics of ice and subglacial meltwater. In this paper we analyse the properties of ice streams that form in a coupled model of ice flow and subglacial hydrology. We see that there is a natural length scale defining ice stream separation and width. This arises as a result of the balance between effective pressure gradients driving meltwater away from ice streams and the enhanced water production in the streams due to the fast ice flow. We further discuss how the model interacts with topography and we show that small perturbations to a uniform bed have a strong effect on where ice streams emerge in the model. However, in many cases ice streams then evolve to be closer to the dimensions defined by the natural length scale of the unperturbed system. The non-dimensional parameter that defines this length scale is therefore of fundamental importance in the model.
\end{abstract}

%
%

%


%
%

\section{Introduction}

Ice streams account for up to 90\% of the Antarctic ice flux into ice shelves and ultimately into the sea \citep{Rignot:2011ko,Bamber:2000dv}. There are many potential mechanisms controlling their formation and subsequent evolution \citep[e.g.\,][]{Langley:2014jg,Winsborrow:2010vp}; to understand past and future behaviour of ice sheets it is important to develop quantitative models and use these to explore the sensitivity of ice-stream behaviour to environmental and basal controls. 

For some ice streams, the mechanism leading to their formation is obvious; large variations in bedrock topography cause ice flow to be channelised into bedrock troughs. The thicker ice in the trough then leads to larger shear stresses and hence faster flow (e.g.\,Rutford Ice Stream, West Antarctica \citep{Doake:2001ri} and Jakobshavn Isbrae, Greenland \citep{CLARKE:1996tz}). However in other cases, topography does not appear to be the main control on size and location of the ice streams. The most prominent example of such `pure' ice streams today are the those found on the Siple Coast region of West Antarctica \citep[e.g.\,][]{Shabtaie:1987kp, Paterson:1994,Bindschadler:2001uy}. 

For at least two decades scientists have known the potential importance of the subglacial hydrologic system in allowing large variations in ice velocities \cite[e.g.\,][]{Alley:1991aa,Engelhardt:1997ut,Bentley:1998ec,Kamb:2001ur}. Rapid velocities are associated with the presence of basal meltwater, and deformable, wet sediment slurries, which allow the ice to slide with less internal deformation \cite[e.g.\,][]{Iken:1986wr,Alley:1986aa,Blankenship:1986aa,Clarke:2005wj}. The behaviour of meltwater at the base of an ice sheet could therefore play a key role in governing temporal and spatial variation in ice velocities; this idea is supported by growing evidence for an active subglacial water system across Antarctica \citep[e.g.\,][]{Fricker:2007ec,Bell:2008aa,Carter:2013aa}. In particular, the layer of till found at the bottom of ice streams is thought to be water-saturated and to deform with Coulomb-plastic behaviour \citep{Iverson:1998uf,Tulaczyk:2000ug,Kamb:2001ur}. Many models of ice-streaming regions, however, lack any detailed description of water transport over the till \citep[e.g.\,][]{Bougamont:2011go}. In previous work, motivated by such evidence and observations, the present authors considered a coupled model of ice dynamics and subglacial meltwater that reproduced the sharp changes in ice velocity observed between ice-stream and inter-ice-stream regions \citep{KyrkeSmith:2013gv}. We solved a numerical model of the fully coupled ice-water system, assuming the meltwater flows in a Weertman-style film \citep{Weertman:1972tg} over saturated, relatively impermeable till, with subglacial, subgrid-scale roughness parameterised in the manner of \cite{Creyts:2009ho}. Solutions showed that in a specific regime, feedbacks arising from the coupling of the meltwater and ice result in emergence of ice streams in the model. 

There are, however, many issues left unaddressed by the previous work, including an analysis of what physical mechanisms govern the size and spacing of ice streams that emerge in the model, and how this may correspond with what is observed in nature. In this manuscript we therefore expand our analysis of ice streams that appear as a consequence of coupled ice and meltwater flow. We investigate influences on the length scales that separate modelled ice streams by considering a balance between the forces that drive meltwater towards and away from ice streams. We also consider the interaction of ice and water flow with lateral topographic variation. A clear distinction is made in the literature between `pure' ice streams and `topographic' ice streams; we quantify, in the context of this model, the amplitude of topography required for bedrock variation to become the main control on the lateral length-scales of ice streams.

The paper is organised as follows. Section 2 is a description of the model. Section 3 addresses the controls on the spacing between ice streams that form over a flat bed. Section 4 considers the influence of topography on the formation, location and spacing of ice streams. We make some concluding remarks in Section 5. 

\section{Model Description}
\label{sec:model}

We provide an outline of the ice and subglacial water flow models below. We present the equations in dimensionless form; the dimensional model is presented in \cite{KyrkeSmith:2013gv}, and from here we guess approximate scales for some variables based on observations/physical intuition, and then use these to derive characteristic scales for other variables, as given in table \ref{tab:scalings}. 

\subsection{Ice Flow}
To determine ice thickness and velocity we use a model that is a vertically-integrated hybrid of the shallow-ice approximation (SIA) and shallow-shelf approximation (SSA). It takes into account both vertical shear stresses and membrane stresses, so providing a unified flow description for all flow regimes within a shallow ice sheet. The force balance includes basal stresses, driving stresses and membrane stresses. A complete description of the model is presented in \citet{KyrkeSmith:2013hk}, and a statement of the non-dimensional equations is given below.

Mass conservation and force balance for the ice are given by
\begin{align}
\label{eq:masscons}
\frac{\partial h}{\partial t} &+ \boldsymbol{\nabla}\cdot\left(h\boldsymbol{\mathrm{u}}_b - \lambda \frac{h^{n+2}}{n+2}\left|\boldsymbol{\gamma}\right|^{n-1}\boldsymbol{\nabla}s_i\right)=a, \\
\boldsymbol{\tau}_b &= h\boldsymbol{\gamma},  \label{eq:taugen}
\end{align}
where $h$ is ice thickness, $\boldsymbol{\mathrm{u}}_b = \boldsymbol{\mathrm{u}}(b)$ is basal ice velocity, $s_i$ is ice surface elevation, $\boldsymbol{\tau}_b$ is basal stress, $a$ is the accumulation rate and
\begin{equation}
\label{eq:sfgamma}
\boldsymbol{\gamma} = -\boldsymbol{\nabla}s_i + \frac{\varepsilon^2}{\lambda}\boldsymbol{\nabla}\cdot\boldsymbol{\mathrm{S}},
\end{equation}
where $\boldsymbol{\mathrm{S}}$ is the resistive stress tensor \citep{VanderVeen:1999,Hindmarsh:2012ba}, 
\begin{equation}
\boldsymbol{\mathrm{S}} = \boldsymbol{\tau} + \boldsymbol{\mathrm{I}}\,\text{trace}\left(\boldsymbol{\tau}\right) =  
\begin{bmatrix}
       2\tau_{11} + \tau_{22} & \tau_{12}       \\
       \tau_{12}            & \tau_{11} + 2\tau_{22} 
     \end{bmatrix}.
\end{equation}
All variables and their scales are given in Table \ref{tab:scalings} and definitions of non-dimensional parameters are given in Table \ref{tab:nondimconstants}.

\subsection{Subglacial water flow}
\label{subsec:massconswater}

\begin{figure}[htb]
\begin{center}
\includegraphics[width=0.95\textwidth]{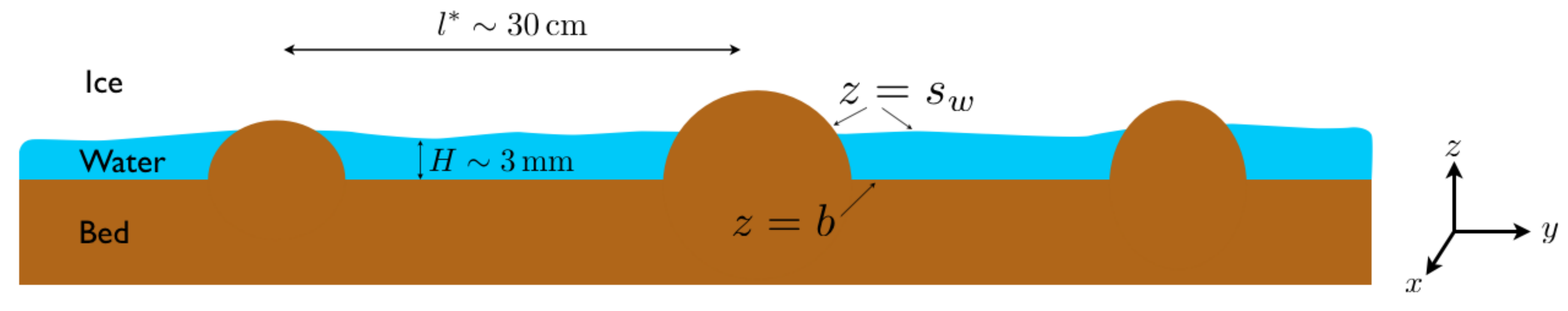}
\caption{Geometry of the rough-bedded, thin-film subglacial water flow configuration}
\label{fig:bedsetup}
\end{center}
\end{figure}

We consider water flowing at the base of an ice stream in a rough-bedded-thin-film configuration as illustrated in Figure~\ref{fig:bedsetup}. The ice is underlain by sediments that are saturated with water; we presume that the till is not very permeable and expect the water at the bed to flow in some kind of distributed system. We therefore consider a Weertman-style film \citep{Weertman:1972tg} and parameterise the subglacial, subgrid-scale roughness in the manner of \cite{Creyts:2009ho}. 

Given that the water flows in a film of depth $H$, and assuming incompressibility of water, mass conservation of water takes the following dimensionless form,
\begin{equation}
\label{eq:nondimHeqnicers}
\delta_T\frac{\partial H}{\partial t} + \boldsymbol{\nabla}\cdot \boldsymbol{\mathrm{q}} = \Gamma,
\end{equation} 
where  
\begin{equation}
\label{eq:meltrate}
\Gamma =   1 + \mu \boldsymbol{\mathrm{u}}_b\cdot\boldsymbol{\tau}_b - \kappa \left|\boldsymbol{\mathrm{u}}_b\right|^{1/2}
\end{equation}
is the dimensionless melt-rate of the ice (the terms correspond to geothermal heating, frictional heating, and cooling through a thermal boundary later respectively \citep{KyrkeSmith:2013gv}). The dimensionless water flux is
\begin{equation}
\label{eq:waterflux}
\boldsymbol{\mathrm{q}} =   - H^3\left(\boldsymbol{\nabla} s_i + \beta\boldsymbol{\nabla} b -  \nu\boldsymbol{\nabla} N \right),
\end{equation}
driven predominantly by gradients in ice surface elevation $s_i$, with small contributions also from gradients in basal elevation $b$, and effective pressure
\begin{equation}
\label{eq:Nexpression}
N = \left[\Lambda(H)\left(\Gamma -  r\delta_T\frac{\partial H}{\partial t}\right)\right]^{1/n}. 
\end{equation}
This expression for dimensionless effective pressure is derived from a closure relation balancing the viscous creep of ice with the melt-back \citep{KyrkeSmith:2013gv}. In the above $\delta_T, \mu, \kappa, \beta, \nu$ and $r$ are all non-dimensional parameters defined in Table \ref{tab:nondimconstants}. $\Lambda(H)$ is a function that describes how separation between supporting clasts over the bed changes as a function of water depth, $H$. It is related to the length scale $l^*$  (as labelled in Figure \ref{fig:bedsetup}) by
\begin{equation}
\label{eq:legnthscaleLambda}
l^*(H)= \frac{l_0}{\Lambda(H)},
\end{equation}
where $l_0$ is a length scale that represents the typical clast spacing over the bed, and can be chosen to be consistent with observed effective pressures on the Whillans Ice Stream. A detailed discussion of how the function $\Lambda(H)$ is chosen is given in \citet{KyrkeSmith:2013gv} with the conclusion that a function of the form 
\begin{equation}
\label{eq:Lambda}
\Lambda(H) = \Lambda_\infty + \delta\left(1-\Lambda_\infty\right)\ln \left[ 1 + \exp\left(\frac{H_c - H}{\delta H_c}\right) \right]
\end{equation}
is an arbitrary but suitable choice. $H_c$ is a dimensionless critical water depth corresponding to the size of the largest supporting clasts (and so representing the depth of water at which the clasts are all submerged). $\Lambda_\infty$ is a parameter governing the value of the effective pressure once $H>H_c$; $\delta$ describes how quickly the transition to constant $\Lambda$ occurs near $H=H_c$.

Finally it is necessary to prescribe a basal boundary condition, relating the basal shear stress, $\boldsymbol{\tau}_b$, to the hydrology through a non-dimensional sliding law 
\begin{equation}
\label{eq:slidinglaw}
\boldsymbol{\tau}_b =c_0 \left|\boldsymbol{\mathrm{u}}_b\right|^pN^q\frac{\boldsymbol{\mathrm{u}}_b}{\left|\boldsymbol{\mathrm{u}}_b\right|},
\end{equation}
where the exponents $p$ and $q$ are commonly taken as $1/3$ \citep{Budd:1979aa, Bindschadler:1983aa}. A detailed discussion of how this sliding law relates to an analytical triple-valued sliding law is given in \citet{KyrkeSmith:2013gv}.

We now have a complete non-dimensional description of the ice-water system. Note that thermodynamics are only included through prescription of the melt-rate and we do not explicitly include energy conservation in the model. As a result of this, we are assuming that the bed is always at melting point, which is something for reconsideration in future work, given that there have been observations and theoretical studies showing basal freeze-on also occurs in the vicinity of ice streams \citep[e.g.\,][]{Raymond:2000aa,Vogel:2005aa}.

\begin{table}
\begin{center}
\renewcommand{\arraystretch}{1.25}
\begin{tabular}{@{} l l l @{} }
\hline
Symbol					                       & Description 		       				& Dimensional value \\
\hline
$\boldsymbol{\mathrm{x}} = (x,y)$							     & distance				& $500\,\mathrm{km}$ \\
$h$							     & ice thickness							& $1000\,\mathrm{m}$ \\
$b$							     & bed elevation							& $\sim h$ \\
$s_i = b+h$							     & ice surface elevation							& $\sim h$ \\
$H$							     & water layer depth			         & $3\times10^{-3}\,\mathrm{m}$ \\
$N$							     & effective pressure				& $4\times10^{4}\,\mathrm{Pa}$\\
$q$							     & meltwater flux				         & $5\times10^{-5}\,\mathrm{m}^{2}\,\mathrm{s}^{-1}$ \\
$\boldsymbol{\mathrm{u}} = (u,v)$							    & horizontal ice velocity					& $100\,\mathrm{m}\,\mathrm{yr}^{-1}$ \\
$a$							    & accumulation rate				& $0.2\,\mathrm{m}\,\mathrm{yr}^{-1}$ \\
$\Gamma$					& melt rate				& $3\,\mathrm{mm}\,\mathrm{yr}^{-1}$ \\
$t$							     	    & time						& $5000\,\mathrm{yr}$ (ice)\\
							    & 							& $1\,\mathrm{yr}$ (water) \\
$\tau$							    & basal stress 					& $2\times10^4\,\mathrm{Pa}$ \\
$\psi$							    & hydraulic potential			& $10^7\,\mathrm{Pa}$ \\
$\Omega$       					    & ice surface slope				& $2\times10^{-3}$ \\
$l_0$							    & clast spacing						& $0.3\,\mathrm{m}$ \\
$p$							& velocity exponent in sliding law              & $1/3$ \\
$q$							& effective pressure exponent in sliding law  & $1/3$ \\
$c$                                  			& sliding law coefficient                              & $6.8\times10^4\,\mathrm{Pa}^{2/3}\,\mathrm{s}^{2/3}\,\mathrm{m}^{-1/3}$ \\
$n$ 							& Glen's flow law exponent			& 1 \\
$A$							& Glen's flow law rate scale			& $1.25\times10^{-23}\mathrm{s}^{-1}\mathrm{Pa}^{-3}$ \\
\hline
\end{tabular}
\end{center}
\caption{Table of variables with typical dimensional values chosen based on scales of current day ice streams on the Siple Coast \citep{KyrkeSmith:2013gv}.}
\label{tab:scalings}
\end{table}

\begin{table}
\begin{center}
\renewcommand{\arraystretch}{2.0}
\begin{tabular}{@{} l l l @{} }
\hline
Symbol			& Definition		& Typical value\\
\hline
$\lambda$		&$\frac{\tau_0 d_0}{\eta_I u_0}$		& $0.0625$ \\
$\varepsilon $ 		&$\frac{d_0}{x_0}$					& $0.002$ \\
$\delta_T$		&$\frac{t_w}{t_i}$			&$2\times10^{-4}$ \\
$\mu$		&$\frac{\tau_0u_0}{G}$			&$1$ \\
$\kappa$			& $\left(\frac{\rho_i c_p k u_0}{\pi x_0}\right)^{1/2}\frac{\Delta T}{G}$	& $0.27$ \\
$\beta$				&$\frac{\Delta\rho_{wi}}{\rho_i}$				& $0.1$ \\
$\nu$			&$\frac{N_0}{\rho_i g\Omega_0 x_0}$	& $4\times10^{-3}$ \\
$r$				&$\frac{\rho_i}{\rho_w}$				& $0.9$ \\
$c_0$			& $\frac{cu_0^{1/3}N_0^{1/3}}{\tau_0}$						& $1.7$ \\
$\Lambda_\infty$	& 						& $0.01$ \\
$\delta$			& 						& $0.01$ \\
\hline
\end{tabular}
\end{center}
\caption{Dimensionless parameters. Subscripts zero refer to the dimensional value of the corresponding variable, and other parameters are constants using standard notation from \citet{KyrkeSmith:2013gv}.}
\label{tab:nondimconstants}
\end{table}

\section{Ice-stream scales}
\label{sec:spacing}

\citet{KyrkeSmith:2013gv} solve the fully coupled ice-water system \eqref{eq:masscons},\,\eqref{eq:taugen},\,\eqref{eq:nondimHeqnicers}\,\&\,\eqref{eq:slidinglaw} numerically. They show that if there is a sufficiently large amount of meltwater produced (e.g.\,if ice flux is large and therefore frictional heating rates are high), the laterally uniform, coupled subglacial water and ice sheet can become unstable, with spontaneous ice stream formation as a consequence. This instability occurs as more of the bed becomes submerged in water, decreasing the basal resistance on the ice. The ice therefore slides more rapidly. It is lateral shear stresses and the thinning of the ice from increased outflow from the domain that prevent the ice increasing in speed indefinitely \citep{KyrkeSmith:2013gv}. The instability is discussed in some detail in \citet{KyrkeSmith:2013gv}, however that paper provides no detailed study of the properties of the resulting ice streams; we attempt to do this here. 

One of the notable features of the instability is that it emerges on a relatively short wavelength and there is a subsequent coarsening of the emergent, alternating, fast and slow flow lateral pattern as the system evolves. Meltwater is routed towards larger, fast-flowing patches, causing the smaller ones to `switch off' due to a lack of lubrication at the bed. This effect is known as water piracy \citep[e.g.\,][]{Anandakrishnan:1997aa}. For example, over a $500\,\mathrm{km}^2$ domain, $5-6$ thin, fast-flowing streams initially develop (we define a stream as an area of the ice that has a downstream velocity greater than $150\,\mathrm{m}\,\mathrm{yr}^{-1}$). These streams then evolve so that once the system reaches its quasi-steady-state, there are just three distinct ice streams, with a distance of $\sim 60$--$80\,\mathrm{km}$ between streams (the inter-stream region being where the downstream ice velocity $u_b < 150\mathrm{m}\,\mathrm{yr}^{-1}$) \citep[][fig.~6]{KyrkeSmith:2013gv}.  The first problem we address is what determines the width and spacing of the streams. While these two quantities could be very different, our numerical results suggest they are similar, and this is consistent with our theoretical discussion in the appendix. We therefore use the appellations `stream width' and `stream spacing' interchangeably.

Downstream of the ice-stream onset zones, the principal variation in the solution fields is in the cross-stream direction. To understand the lateral spacing between streams we therefore return to the dimensionless mass conservation of water equation \eqref{eq:nondimHeqnicers} and ignore downstream variations and time-dependent terms. We furthermore neglect $\partial s_i/\partial y$, on the physical basis that a significant cross stream slope would be eradicated by the consequent cross stream ice flow, and also on the practical basis that its value is indeed small downstream of the onset zone in numerical simulations (see Figure 7 in the appendix). This leaves the melt-rate term and what we class as a lateral diffusion term that governs the cross-domain flow of meltwater due to gradients in the effective pressure. These two terms must be in balance downstream of the ice stream onset for the system to be in steady-state. Hence we expect
\begin{equation}
\label{eq:crossstreambalance6}
\nu\frac{\partial}{\partial y}\left[H^3\frac{\partial}{\partial y}\left(\Lambda(H)\Gamma \right)^{1/n} \right] \sim \Gamma,
\end{equation}
(c.\,f.\,eqn.\,\ref{eq:nondimHeqnicers}). The lateral spacing between streams is on some length-scale, $y\sim l_y$. Given that the non-dimensional parameter $\nu\ll 1$ and all variables are non-dimensional, for the system to be in balance we therefore require $\nu / l_y^2 \sim 1$. The lateral length-scale $l_y$ consequently scales as 
\begin{equation}
\label{eq:spacingscaling6}
l_y^2\sim\nu = 4\times10^{-3}.
\end{equation}
This is analogous to the half-width of the spacing between streams scaling as $\sim \sqrt\nu$, or in dimensional terms $\sim \sqrt\nu x_0$. In the case where $\nu = 4\times10^{-3}$, as is the estimated value provided in Table \ref{tab:nondimconstants}, this gives a spacing of $64\,\mathrm{km}$, which is roughly in line with that in simulations \citep[][fig.\,6]{KyrkeSmith:2013gv}. Further detail of this discussion is given in the appendix.

The non-dimensional parameter $\nu$ represents the ratio of the effective pressure and hydrostatic ice pressure contributions to the gradients in hydraulic potential, i.e.~
\begin{equation}
\label{eq:wordsnu}
\nu  = \frac{[\boldsymbol{\nabla}N]}{[\rho_i g \boldsymbol{\nabla} s_i]}\sim\frac{N_0}{\rho_i g h_0} ,
\end{equation}
where square brackets are used to represent the scale of each term. The effective pressure contribution to the hydraulic potential gradient drives water away from ice streams. This is because there is the most water at ice stream centerlines (the ice is flowing fastest there, resulting in high frictional heating) and water flux is in the direction of increasing effective pressures, corresponding to decreasing water pressure/depth. In contrast, gradients in hydrostatic pressure drive water toward ice streams; high ice flux causes depressions in surface elevation to develop at their onset zones, which drives meltwater towards the ice stream. The ratio of the magnitudes of these two terms is directly linked to the lateral length scale of the ice-stream arrangement across the domain.

As this ratio increases, the spacing between streams widens. The physical rationale for this is as follows. A larger value of $\nu$ in the model means that the effective pressure gradients play an increased role in governing water flow. In turn this means that a smaller gradient in effective pressure between adjacent streams is required for the system to be in balance (i.e.~to satisfy \eqref{eq:crossstreambalance6}). Now note that $\delta N$---the difference between $N$ at the centerline of the stream and along the inter-stream ridge--- is approximately constant in all cases where streams develop in the model; the choice of the parameter $\Lambda_\infty$ in the function \eqref{eq:Lambda} governs what the effective pressure will be inside the stream where $H>H_c$, and along the inter-stream ridge water pressure $\sim 0$ so the effective pressure is equal to the ice overburden pressure. Crucially, if a smaller gradient is then required for the system to be in balance, this must correspond to a wider spacing, whereby $N$ changes over a larger distance. It is therefore the stream spacing that adjusts for a larger value of $\nu$ as the effective pressure gradients have a more dominant role in governing ice flow.

\begin{figure}
\begin{center}
\includegraphics[width=0.75\textwidth]{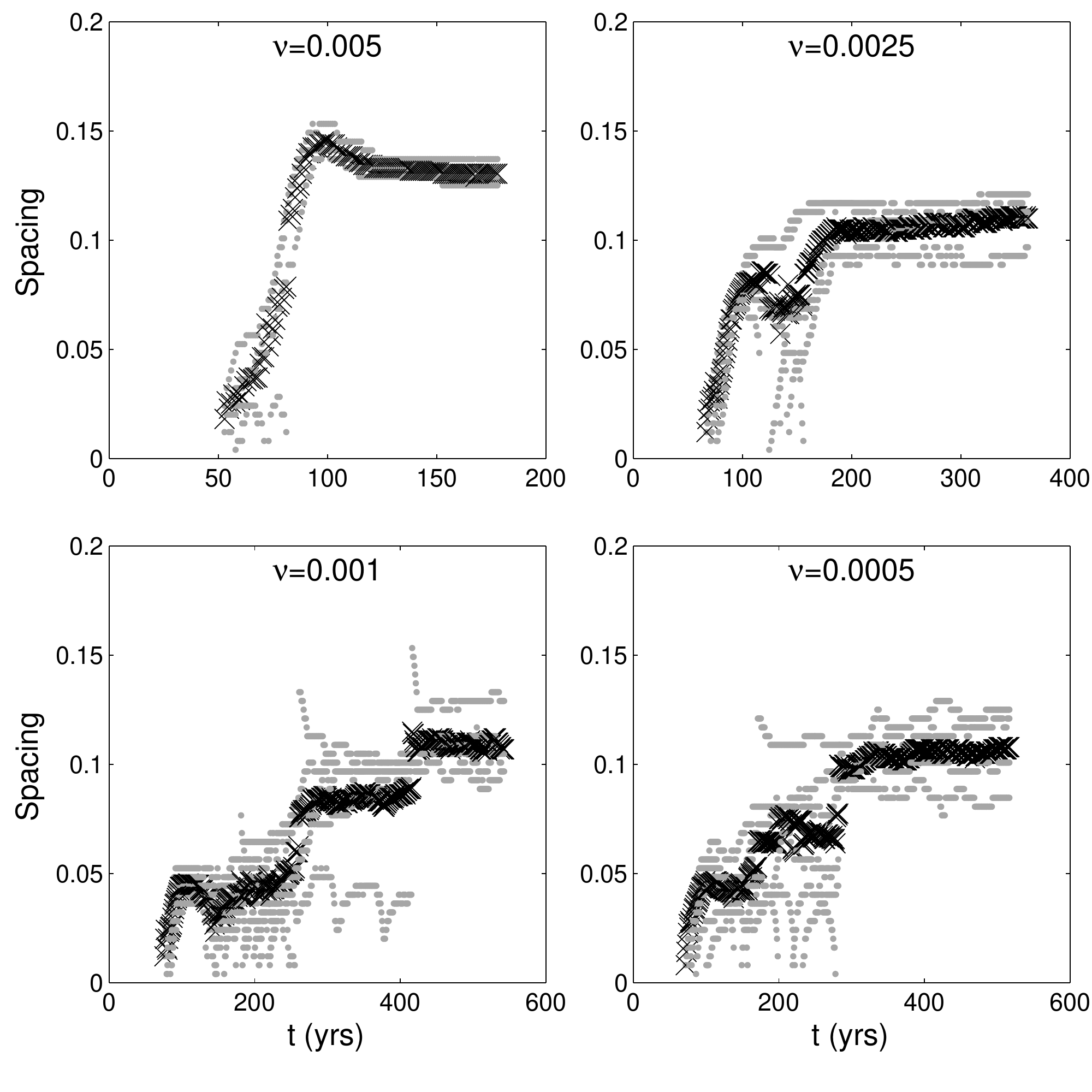}
\caption[Plots of the evolution of the spacing between ice streams with different values of the non-dimensional parameter $\nu$.]{The evolution of spacing between streams, for four simulations run with different values of $\nu$. The spacing is measured $10\,\mathrm{km}$ from the end of the domain, and is defined as the distance over which $u_b < 150\mathrm{m}\,\mathrm{yr}^{-1}$ between two ice-stream shear margins. Grey dots represent individual measurements between fast-flowing streams, and the black crosses the mean spacing for each simulation.}
\label{fig:nuevolvplots}
\end{center}
\end{figure}

To test this hypothesis, we ran a suite of simulations with a range of values of $\nu$ specified and measure the variation in separation between streams. Figure \ref{fig:nuevolvplots} shows time series of the separation between fast-flowing ice streams that form in four different simulations with $\nu$ varied over an order of magnitude. We define $u_b \geq 150\,\mathrm{m}\,\mathrm{yr}^{-1}$ to be in an ice stream and $u_b <  150\,\mathrm{m}\,\mathrm{yr}^{-1}$ to be outside of an ice stream; the spacing is therefore the width over which $u_b <  150\,\mathrm{m}\,\mathrm{yr}^{-1}$, i.e.,\,the distance between ice-stream margins of two side-by-side streams. 

Ice streams first develop at a similar time for all values of the parameter $\nu$ shown in the figure. However, for larger $\nu$, there is less variation in spacing and the system reaches a state where ice streams are no longer merging or splitting more rapidly. For $\nu\gtrsim 0.005$ ice stream formation is suppressed (given the other parameter values used in this study). This is a consequence of the more significant role effective pressure gradients play in driving water flow at higher values of $\nu$. Considering eqn.~\eqref{eq:nondimHeqnicers} with water flux given by eqn.~\eqref{eq:waterflux}, the term for the flux driven by gradients in effective pressure, $\nu H^3\nabla N$, is larger and so the time derivative of $H$ is larger. At larger values of $\nu$ the water is therefore driven more rapidly away from initial perturbations, in the direction of increasing effective pressure. Ice evolution has a much longer timescale than hydraulic evolution; this ratio is described by the non-dimensional parameter $\delta_T$. The surface depressions at the onset of ice streams therefore do not develop as rapidly, and so the $H^3\nabla s_i$ term in the water flux expression \eqref{eq:waterflux} does not begin to play a role as immediately as the gradients in effective pressure. At larger values of $\nu$ the lateral spreading of the water therefore smooths out any perturbations in the water layer before the surface elevation lowers sufficiently for the gradients in surface elevation to start to play a significant role in driving water flow in eqn.~\eqref{eq:nondimHeqnicers}. The depressions are responsible for driving water toward the ice streams, therefore driving the ice-stream forming instability.  When $\nu$ is smaller, $H_t$ is correspondingly smaller and surface depressions may form rapidly enough to allow distinct ice-streams to emerge and stabilise. 

\begin{figure}
\begin{center}
\includegraphics[width=0.7\textwidth]{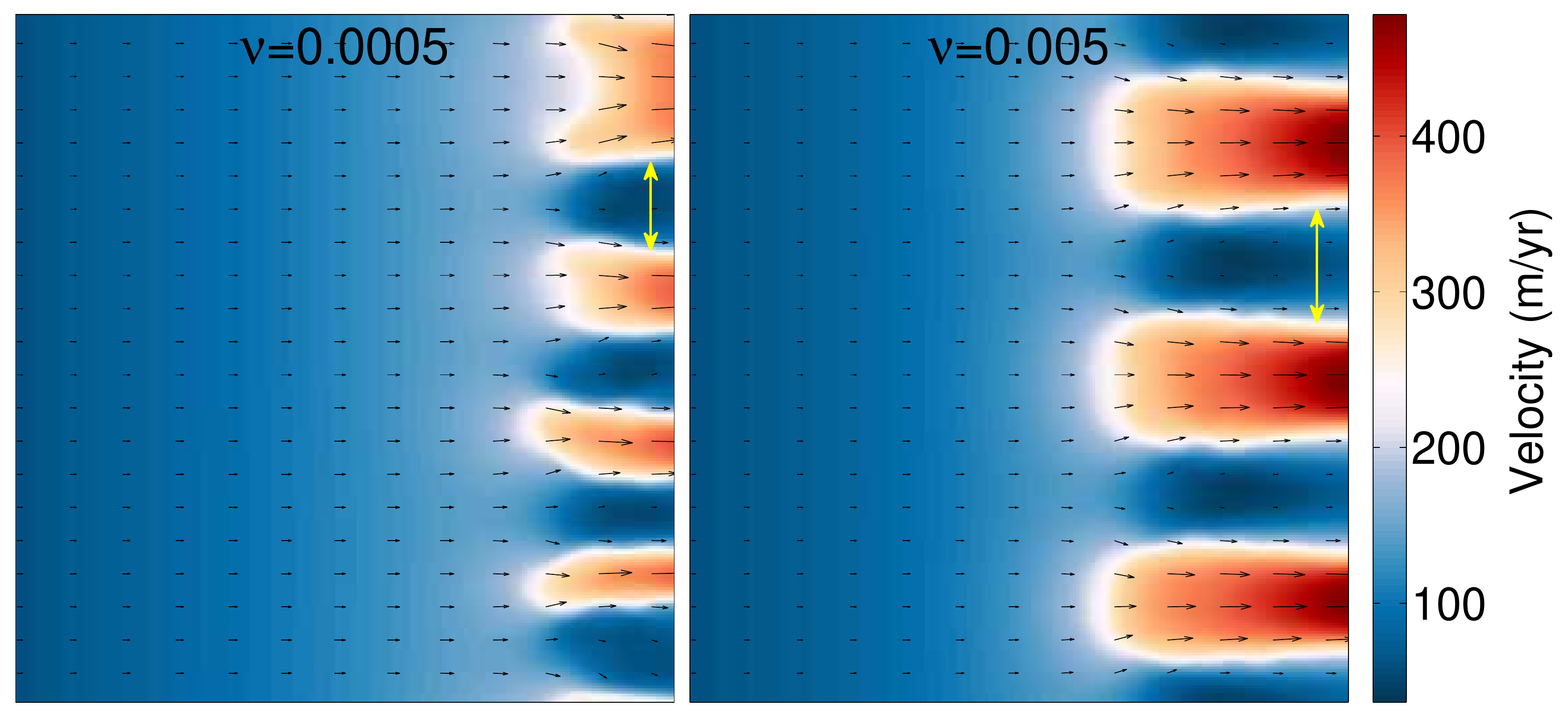}
\caption[Quasi-steady-state solution fields for two simulations with different values of the non-dimensional parameter $\nu$.]{Plots of the velocity field, once it has reached a quasi-steady-state, for two values of $\nu$. The value of $\nu$ affects the spacing of the streams that form. The yellow arrows illustrate what we define as the spacing. The colours represent the magnitude of velocity, measured in $\mathrm{m}\,\mathrm{yr}^{-1}$ and the arrows are ice velocity vectors.}\label{fig:velfieldspacing}
\end{center}
\end{figure}

Two end-member velocity fields are shown in Figure \ref{fig:velfieldspacing}. These are shown at the late stage of quasi-steady-state, when any significant variation in the width and separation of streams has died out. For larger $\nu$, there are fewer ice streams and they are wider, with wider spaces between adjacent streams, as expected from \eqref{eq:spacingscaling6}.

\begin{figure}
\begin{center}
\includegraphics[width=0.65\textwidth]{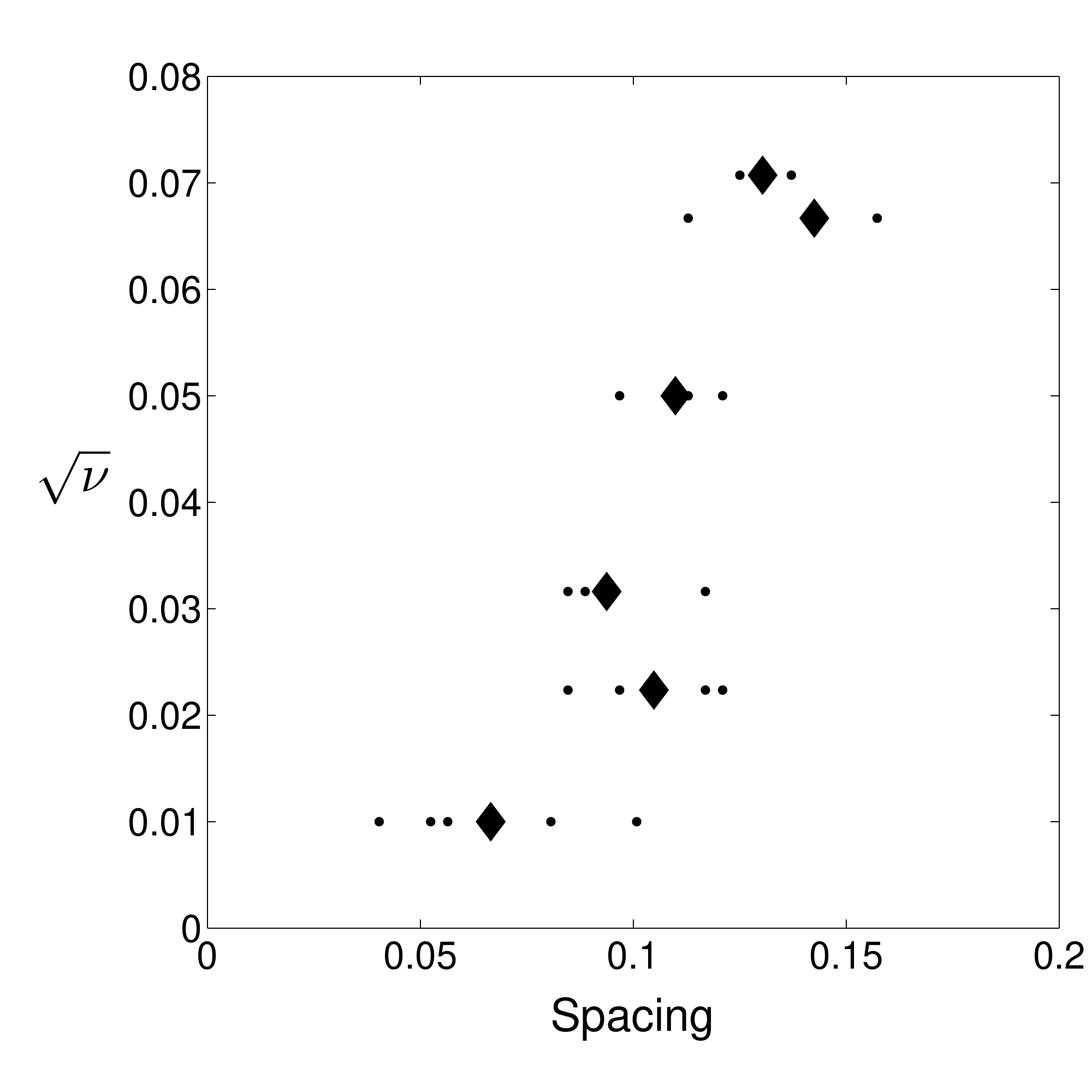}
\caption[Plots of the evolution of the spacing between ice streams with different values of the non-dimensional parameter $\nu$.]{The evolution of spacing between streams, for four simulations run with different values of $\nu$. The spacing is measured $10\,\mathrm{km}$ from the end of the domain, and is defined as the distance over which $u_b < 150\,\mathrm{m}\,\mathrm{yr}^{-1}$ between two ice stream shear margins. The dots represent individual measurements between side-by-side fast-flowing streams, and the diamonds the mean spacing for each simulation.}
\label{fig:spacingnu}
\end{center}
\end{figure}

To relate the spacing to $\nu$ more directly, in Figure \ref{fig:spacingnu} we plot the inter-stream spacing of the streams against $\sqrt{\nu}$. There is a positive correlation, as expected, but a line of the spacing $l_y\propto\nu$ is not a good match to the data. There a number of plausible reasons for this, which include the possibility that the asymptotic limit is not reached at the higher values of $\sqrt{\nu}$, or more simply that our rather noisy definition of stream spacing is not necessarily the most suitable. Furthermore, it is important to note the large scatter in values of the spacing (each measurement is plotted with a small dot). To analyse this variation in detail would require more detailed study, including details of the stress-balances in the ice that are necessary for steady-state ice-stream flow.

The analysis in this section nevertheless shows us that the relative importance of effective pressure and ice surface slope to controlling water flux influences the lateral spacing of ice streams that form in our model. Different bed properties correspond to different values of the non-dimensional parameter $\nu$, since the magnitude of effective pressure varies depending on bed properties and overburden pressures. We therefore expect the lateral length scales of ice streams to vary in different settings.

\section{The influence of topography}

All ice-stream formation discussed up to this point and in \cite{KyrkeSmith:2013gv} occurs without any imposed perturbation to the bed--ice--water system. The ice and water are both initialised with no cross-domain flow, and the bed is also uniform with a constant downstream slope and no cross-flow variation. After the instability develops, ice streams evolve to be spaced apart by a length scale that is dependent on parameters in the problem, as discussed in the previous section. These modelled ice streams are analogous to pure ice streams, where topography plays no role at all in governing where they form.

We therefore ask what role variation in basal topography would play in governing where and when ice streams form in our coupled ice--water system? What level of topographic variation is needed for the pure ice streams to become topographically controlled ice streams, constrained by the variations in bed elevation?

To investigate this problem, we apply a perturbation to  bed elevation in the form
\begin{equation}
\label{eq:bedpert6}
\hat{b} = B_0 \sin\left(2\pi n_b y\right),
\end{equation} 
where $B_0$ is the dimensionless amplitude of the perturbation ($B=B_0\times10^3\,\mathrm{m}$) and $n_b$ the number of oscillations imposed across the domain. The dimensionless wavelength of the perturbations is therefore $1/n_b$. This choice of a cross-stream oscillatory perturbation is motivated by the evidence for glacial and meltwater erosion occurring beneath ice streams \citep[e.g.\,][]{Anderson:2008aa,Wellner:2001aa,Bougamont:2003aa}; we expect that such erosion may be responsible for sub-ice-stream troughs. It is worth noting that this raises a question of causality since the order in which ice streams and troughs appear is debatable. Regardless, testing the effect that topographic variation may have on where streams form can give an idea of how much spatial variation of ice streams can be expected, even in the presence of bedrock troughs (whether or not these have been eroded by already-present ice streams). 
\begin{figure}
\begin{center}
\includegraphics[width=0.9\textwidth]{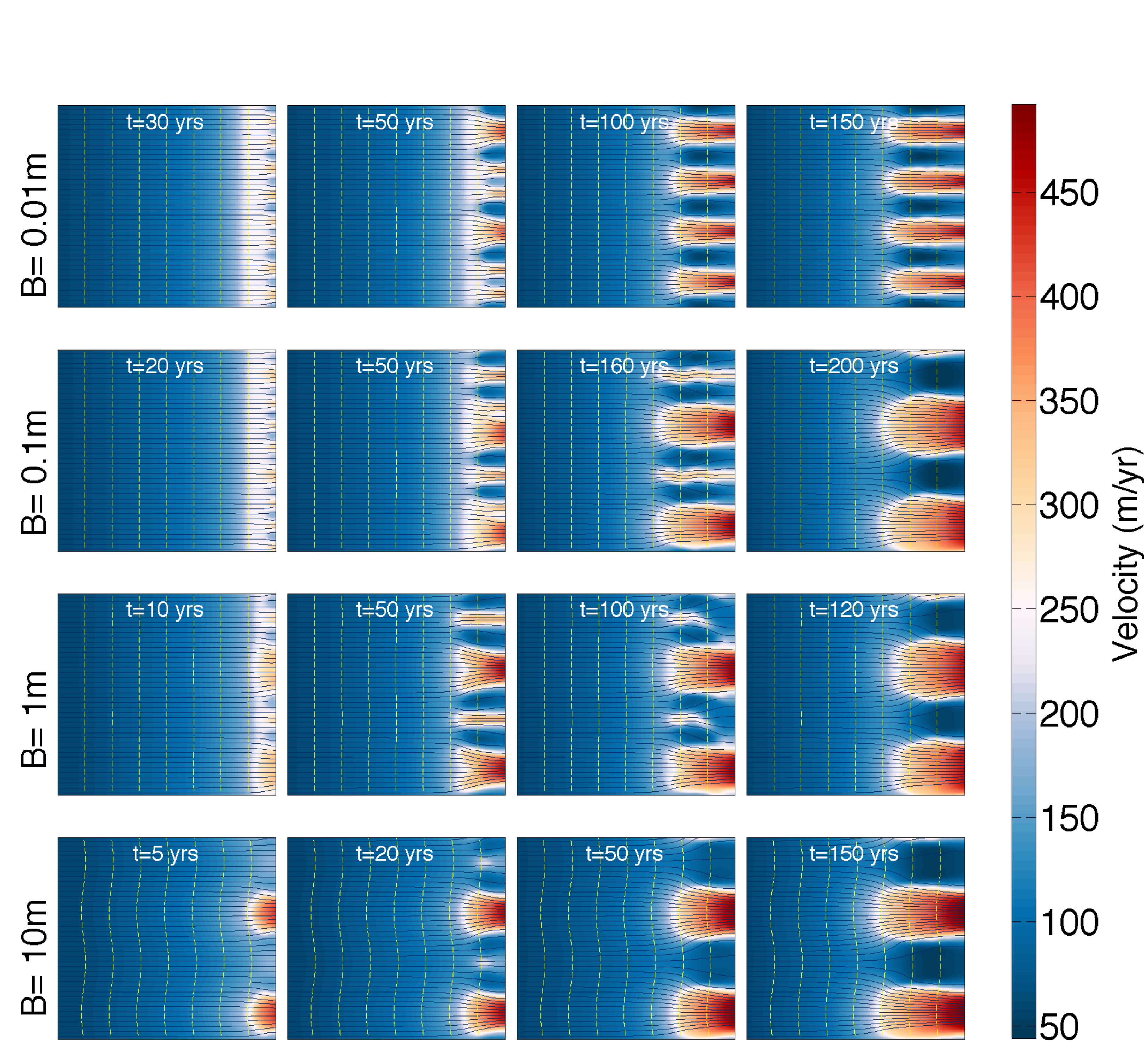}
\caption[Solution fields with a range of amplitudes of bed perturbation.]{Results from a suite of simulations with an applied perturbation in the bed elevation across the domain. The perturbation is of the form \eqref{eq:bedpert6}, with $n_b = 2$ and magnitude $B = B_0\times10^3\,\mathrm{m}$. Each row shows plots at successive times for a simulation with the magnitude of bed perturbation specified (largest perturbation shown in the bottom row of plots). Colours represent ice speed (plotted in $\mathrm{m}\,\mathrm{yr}^{-1}$); solid navy lines are streamlines of water flow in the subglacial hydrologic system; broken yellow lines are contours of basal elevation.}
\label{fig:bedpertfigs}
\end{center}
\end{figure}

Reconsidering the steady-state lateral balance \eqref{eq:crossstreambalance6} in the presence of bedrock variation, we have a third term that becomes important. This is the water flux term due to gradients in bedrock in \eqref{eq:waterflux},
\begin{equation}
\label{eq:bedrockterm}
\beta\frac{\partial}{\partial y}\left[H^3\frac{\partial b}{\partial y}\right] \sim 4\pi^2\beta B_0n_b^2
\end{equation}
with a bed perturbation of the form \eqref{eq:bedpert6}. We therefore expect that if $4\pi^2\beta B_0n_b^2 <<\nu$ then the topography will not provide a strong constraint compared with the effective pressure contribution to water flux; ice streams will behave as in the previous section. In contrast, when 
\begin{equation}
\label{eq:NBbalance}
4\pi^2\beta B_0 n_b^2\gtrsim \nu
\end{equation}
the behaviour will be more strongly influenced by topography. We test this influence by running a set of simulations with an imposed bedrock perturbation of the form \eqref{eq:bedpert6}.

We first consider the behaviour resulting from simulations run with $n_b=2$ (i.e.~two bedrock troughs across the domain). Figure \ref{fig:bedpertfigs} shows snapshots of solutions, plotted at successive times, for four distinct simulations with the oscillation amplitude varied over four orders of magnitude. The times at which plots are shown illustrate the initial onset of the unstable behaviour (first column), the development of distinct fast flowing streams that in some cases merge together (middle columns), and then the final quasi-steady-state (last column). At this wavelength of perturbation, all but the smallest amplitude evolve to have two wide and fast-flowing streams in in the bedrock troughs. For the smallest amplitude perturbation, $B=10^{-2}\,\mathrm{m}$, there are also streams overlying the bedrock peaks. Streams over bedrock highs are also seen (first row of Figure~\ref{fig:bedpertfigs}) in the simulations with larger bed perturbations, but they are not maintained as the system evolves. The thinner streams located between the topographic lows disappear as the streams lying in the troughs widen. As these streams widen, the catchment area for water routing towards the troughs increases, limiting the water supply to the thinner streams located on the peaks of the bed perturbations. There is therefore insufficient water to maintain fast flow in the streams located at the peaks of the perturbations. 

This behaviour is a consequence of the topography providing an additional control on water routing. The leading-order contribution to water flux is from the hydrostatic ice pressure, which causes water to flow in the direction of decreasing surface elevation. However, there are two smaller terms in the expression for water flux \eqref{eq:waterflux} that describe the contributions from gradients in bed elevation and effective pressure. While the contribution from basal gradients is only $1/10$ the size of that from ice surface gradients, this can become non-negligible when basal slopes are $\sim10\times$ as large as surface slopes. The topography becomes a control on water flux, driving water towards bedrock troughs. This effect of the topography on water routing also explains the variation in time scales seen between simulations. The time over which the water piracy occurs (resulting in the streams at bedrock peaks switching off) increases as the the bed perturbation decreases in amplitude. This not only explains why the initial instability develops more rapidly when the bed perturbation is larger, but also reveals why streams located outside of the troughs shut down earlier in the simulation when the topography has larger amplitude. At larger amplitude, water routing towards the topographic lows is stronger. In the case of the largest perturbation ($B = 10\,\mathrm{m}$), the presence of fast-flowing patches at the bedrock peaks is not maintained beyond $t=20\,\mathrm{yrs}$. In contrast, the system reaches its two-stream, quasi-steady-state at around  $t=100\,\mathrm{yrs}$ and $t=160\,\mathrm{yrs}$ for $B = 1\,\mathrm{m}$ and $B=0.1\,\mathrm{m}$ respectively.

Furthermore, the topography variations also modify the flow of ice. This provides a secondary feedback, where the ice is channelised into troughs, lowering the surface elevation above the troughs and therefore providing a stronger control on the water routing towards them.

\begin{table}
  \centering
  \renewcommand{\arraystretch}{2.0}
  \setlength{\tabcolsep}{16pt}
  \begin{tabular}{@{}r|ccccc@{}} 
     & \multicolumn{5}{c}{$\bf{\it{n_b}}$} \\
     $\bf\it{{B}}$ &\bf{1}&\bf{2}&\bf{3}&\bf{4}&\bf{5} \\
    \hline{}
\bf{100\,m} &\checkmark\,\,\,\checkmark &\checkmark\,\,\,\checkmark&\checkmark\,\,\,\checkmark &\checkmark\,\,\,\checkmark & \checkmark\,\,\,\checkmark\\
    \bf{10\,m} &$\times$ \checkmark &\checkmark\,\,\,\checkmark &\checkmark\,\,\,\checkmark &\checkmark\,\,\,\checkmark& \checkmark\,\,\,\checkmark\\
  \bf{1\,m} &$\times$\,\,\,$\times$&$\times$\,\,\,\checkmark  &\checkmark\,\,\,\checkmark &\checkmark\,\,\,\checkmark & \checkmark\,\,\,$\times$ \\ 
     \bf{0.1\,m} &$\times$\,\,\,$\times$& $\times$\,\,\,\checkmark & \checkmark\,\,\,\checkmark &\checkmark\,\,\,\checkmark & \checkmark\,\,\,$\times$ \\
     \bf{0.01\,m} &$\times$\,\,\,$\times$&$\times$\,\,\,$\times$&$\times$\,\,\,\checkmark &$\times$\,\,\,\checkmark & \checkmark\,\,\,$\times$\\
     \end{tabular}
  \caption[Results for how the frequency and amplitude of implemented topography affects where ice streams form.]{Observations of whether streams form in troughs (tick) or have positioning independent of topography (cross). The first entry for each case refers to whether streams initially develop only in the bedrock troughs and the second whether streams remain only in troughs. $B$ is the dimensional amplitude of bed oscillation $(B = B_0\times10^3\,\mathrm{m})$.}
  \label{tab:bedperts}
\end{table}

Having considered variations in amplitude of a cross-stream bedrock perturbation, we extend the analysis to different wavelength perturbations. How does the number of cross-stream oscillations in topography change the behaviour and the positioning of ice streams that form? Table \ref{tab:bedperts} summarises results from simulations run with a range of values of $B_0$ and $n_b$.  A tick mark means that the topography variations govern ice-stream placement (i.e.~the fast flowing patches of ice immediately form only in the bedrock troughs), whereas a cross corresponds to ice-stream features forming independently of the topography (i.e.~fast flowing patches form, at least temporarily, outside of the bedrock troughs). There are two entries for each simulation---the first refers to the initial formation of the ice streams, and the second to the point at which the simulation has reached its quasi-steady-state. It is evident that the longest wavelength of topography perturbation requires the largest amplitude perturbation to have ice streams only in the troughs. More specifically, in the case of there being one bedrock trough across the domain, only with the largest of perturbations ($B=100\,\mathrm{m}$) does just one ice stream develop in the domain, in the bedrock trough. With any smaller perturbation, while a larger ice stream develops in the trough, streams do still form in the other half of the domain. A larger wavelength means the change in basal elevation is taking place over a larger length scale, corresponding to smaller basal gradients and hence the topography provides a lesser control on water flux (by eqn.\,\ref{eq:waterflux}). 

In contrast, in the case when there are three or four bedrock peaks and troughs cross-domain ($n_b = 3,4$), the streams always form in the bedrock troughs, even with a perturbation as small as $B = 1\,\mathrm{cm}$. This is not entirely surprising, since the pattern of the pure streams on a flat bed reaches a state with approximately 3--4 ice streams across the domain. However, as the wavelength of the perturbation decreases further, the behaviour changes again. Initially streams form in the troughs, since a smaller wavelength corresponds more closely to the natural wavelength at onset of the instability. As the system evolves, the pattern coarsens and streams span several troughs, transitioning back towards the natural length scale associated with streams on a flat bed. This is a consequence of the water being driven away from beneath the ice streams by the gradients in effective pressure. When bed perturbations are small, this effect is comparatively large enough to drive water away from the trough, and streams therefore merge and the pattern coarsens as observed on a flat bed. With larger bedrock variations, the gradients in effective pressure are not large enough to have this effect because the gradients in both surface and bedrock elevation driving water towards the ice streams provide stronger controls (in agreement with eqn.~\eqref{eq:NBbalance}).

\begin{figure}
\begin{center}
\includegraphics[width=0.9\textwidth]{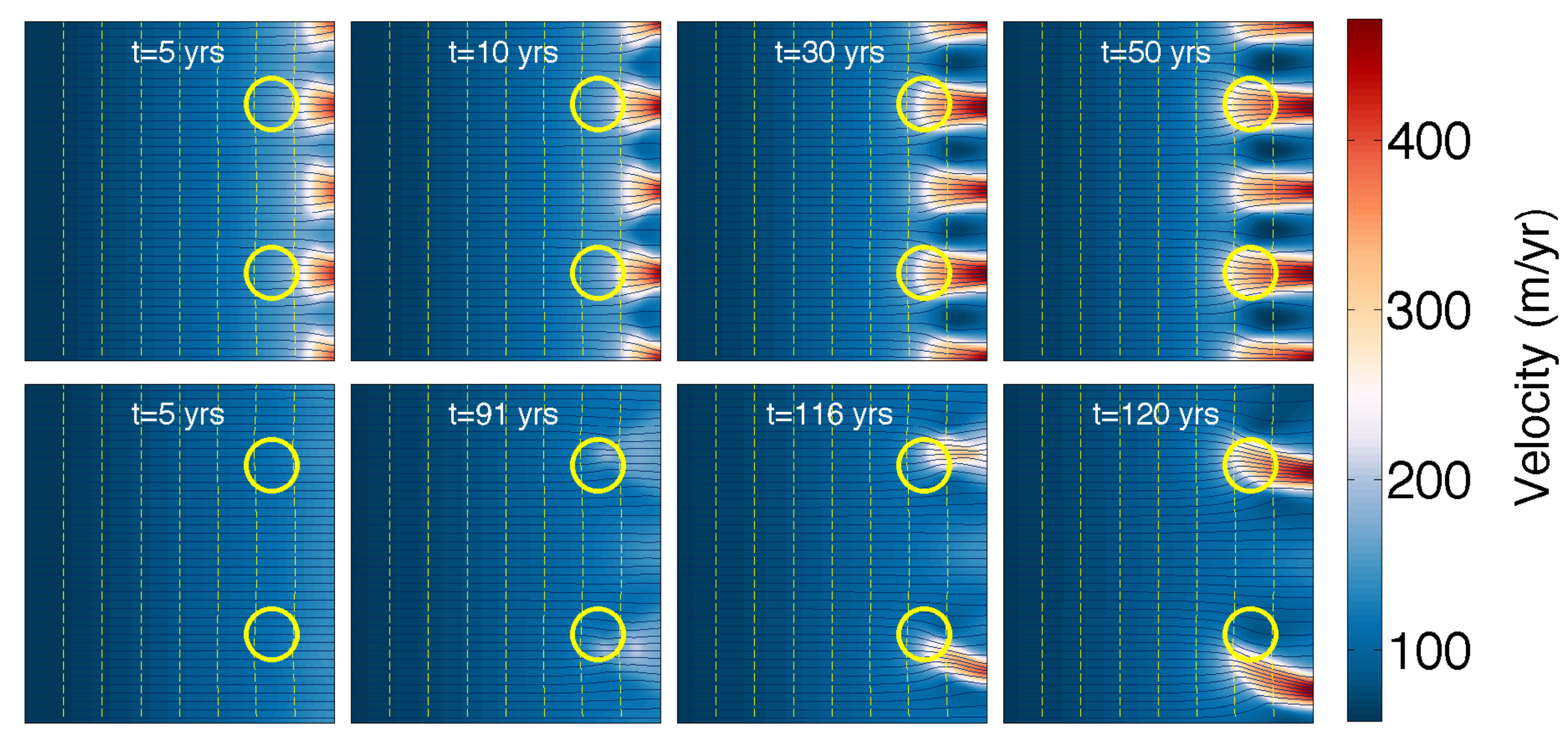}
\caption[Solution fields with Gaussian perturbations]{Results from two simulations with an applied perturbation in the bed elevation in the form of two Gaussian depressions in bed elevation. The plots in the first row are from a simulation run with a critical water depth $h_c = 1.4$, as for all other simulations in this paper, and the second row are from a simulation run with $h_c = 1.8$. Each row shows plots at successive times. Colours represent ice speed (plotted in $\mathrm{m}\,\mathrm{yr}^{-1}$); solid navy lines are streamlines of water flow in the subglacial hydrologic system; broken yellow lines are contours of basal elevation; solid yellow circles illustrate where the Gaussian depressions are centred. }
\label{fig:bedpertdimplefigs}
\end{center}
\end{figure}

As a final experiment, we consider perturbations applied to the bedrock that are not uniform in the downstream direction. While observations of troughs beneath ice streams motivated the last choice of perturbation, there is a growing number of observations of topographic lows at the onset zones of streams \citep[e.g.\,][]{Blankenship:2001ur}, many in the form of subglacial lakes \citep[e.g.\,][]{Bell:2007aa}. For this experiment we therefore impose two Gaussian depressions in bed elevation, $10\,\mathrm{m}$ in amplitude, centred on non-dimensional coordinates $(0.8, 0.25)$ and $(0.8, 0.75)$, on top of the uniform linear bedslope from Section \ref{sec:spacing}. Figure \ref{fig:bedpertdimplefigs} shows snapshots of the solution for two different simulations. The top row solutions are from a simulation run with $h_c = 1.4$ (a value of the critical water depth that gives instability in the case of a flat bed \citep{KyrkeSmith:2013gv}), and the bottom row takes a deeper critical water depth, $h_c = 1.8$. In the case with $h_c = 1.4$, streams form as expected, however the presence of the dimples in the bed regularises the locations, and one can see that the resulting streams are slightly larger and faster flowing when their onset is in the region of a bed depression. In the case where $h_c = 1.8$, streams form only downstream of the depressions in the bedrock elevation. This is because water converges towards the depressions (eq.~\ref{eq:waterflux}) and only then accumulates enough to reach the critical water depth and hence initiates fast flow downstream. With this value of $h_c$ on a flat bed, streams do not form because the water never reaches the critical depth. Therefore, while the critical water depth $h_c$ is narrowly constrained for a perfectly flat bed \citep{KyrkeSmith:2013gv}, it can take a broader range of values and still produce ice streams when the bed isn't perfectly flat---and this corresponds more closely to reality. More specifically, stream formation in the presence of bedrock perturbations is a consequence of whether the gradients in bedrock elevation are steep enough to deflect water towards the depressions, such that the critical depth is reached. This example therefore illustrates a mechanism of stream formation where water has built up in the vicinity of bedrock lows.

\section{Summary and conclusions}
\label{sec:conclusions}

In this paper we have quantified the development of ice-stream-like features that form as a result of the coupled interactions between ice and subglacial water flow. Most notably, we have derived an expression for the horizontal, cross-flow lengthscales associated with ice stream width and separation and we have related this to the balance between effective pressure  gradients driving water away from ice streams and the enhanced rate at which this water is generated within these streams by the fast ice flow. This is an intriguing result because the magnitude of the effective pressure is directly linked to subglacial bed conditions. While parameter values used in \citet{KyrkeSmith:2013gv} were chosen to be consistent with Siple Coast observations and resulting ice streams from the model were of similar scale and magnitude to those observed, in this paper we have varied the value of the non-dimensional parameter $\nu$ and seen that resulting ice streams have different spatial properties. We therefore suggest that this relationship between the horizontal length scale and the parameter $\nu$ can provide a useful comparison point for data from paleo-ice-streams. There is much discussion in the literature about properties of the bed beneath paleo-ice streams \citep[e.g.\,][]{Cofaigh:2002aa,Dodswell:2004aa,Piotrowski:2001va,Stokes:2003fo}. Given the present results, it seems plausible that the location and dimensions of paleo-ice streams \citep[e.g.\,][]{Stokes:2001aa} can be used to estimate effective pressures underneath the ice streams, therefore providing some constraint on the hydrologic system. This could provide an interesting path of investigation for future work. 

Furthermore, we have considered the effect of introducing bedrock variation into the model. For wavelengths of topographic variation that are similar to the natural length scale of the pure ice-stream pattern on a flat bed, only very small magnitudes of topography are required to influence where ice streams form. However, for larger or smaller wavelength oscillations, ice streams do not remain constrained by the topography. At large-wavelength bedrock oscillations, the perturbation needs to be $\mathcal{O}(100\,\mathrm{m})$ for ice streams to form only in bedrock troughs. This is a consequence of the fact that to leading order, the water flux is driven by ice surface slopes; gradients in bedrock elevation only start to play a more significant role when their slope exceeds that of the ice surface by a factor of $\sim 10$. At much smaller wavelengths, even small amplitude bedrock oscillations govern initial placement of ice streams. However, some of the streams then merge and the associated ice-stream length scales become closer to the natural length scale of the pure streams on a flat bed. 

Finally we also considered the effect of perturbations to the bedrock in the form of dimples in the bed. We see that isolated bedrock lows may indeed provide a constraint on where ice streams form, as water flows towards the depressions; a build up of water can result in sufficient lowering of the frictional resistance to allow the onset of ice-stream flow. This is of particular interest given recent observations of subglacial lakes at onset zones of ice streams \citep[e.g.\,][]{Bell:2007aa}. 

In conclusion, while this paper discusses the detailed physical mechanisms that govern the length scales of ice stream separation due to the coupled model dynamics of subglacial hydrology and ice flow, more work is required to compare our results with observations. The present work has provided a step forward in quantifying properties of ice streams by relating the behaviour of the ice with that of the subglacial hydrology, and provides a theoretical basis for future work in the area. 

\appendix

\section*{Appendix}


\begin{figure}[!t]
\begin{center}
\includegraphics[width=0.7\textwidth]{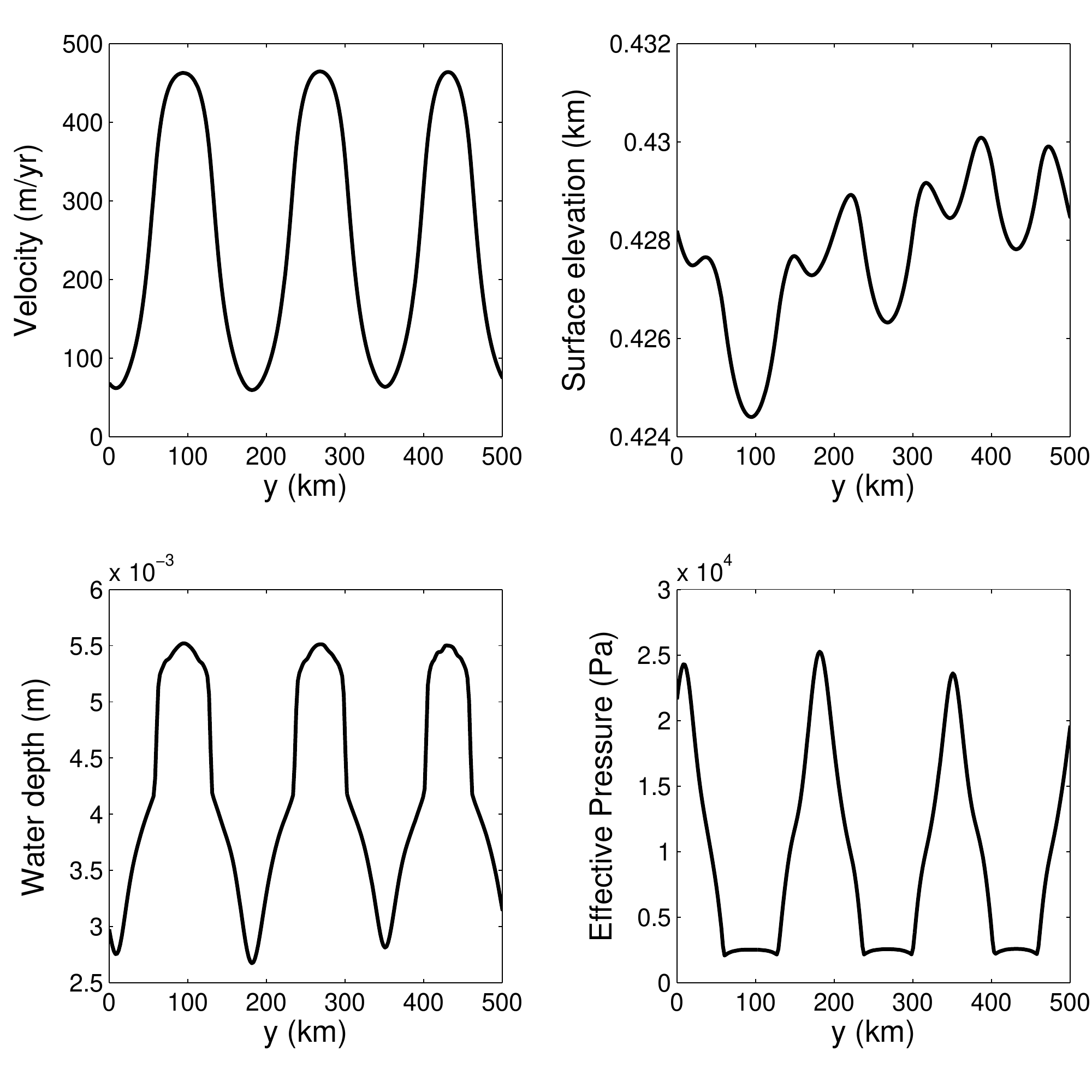}
\caption{\label{figA1}A typical cross section view of water depth $H$, surface elevation $s_i$, downstream velocity $u_b$ and effective pressure $N$ from a computation with $\nu=0.005$ (corresponding to the second velocity field in Figure 3).}
\end{center}
\end{figure}

In order to focus on the issue of stream width, we consider equations \eqref{eq:nondimHeqnicers}, \eqref{eq:meltrate}, \eqref{eq:waterflux}, \eqref{eq:Nexpression} in the main text. Figure \ref{figA1} shows that of the two lateral water flux terms downstream of the onset zone, the surface slope contributes a small additional focussing mechanism, so it is only the $\partial N/\partial y$ term that acts to balance the blow up of $H$, hence the choice of scaling in \eqref{eq:crossstreambalance6}. Similarly, in a steady state over a flat base we take $H$ to satisfy the approximate equation
\begin{equation}
\label{a1}
(H^3)_x=-\nu\frac{\partial}{\partial y}\left[H^3\frac{\partial N}{\partial y}\right]+\Gamma,
\end{equation}
where  in simplistic versions of equations (6), (8) and (11), we take 
\begin{equation}
\label{a2}
\Gamma\sim u\sim\frac{1}{N}\sim\frac{1}{[\Gamma \Lambda(H)]^{1/n}},
\end{equation}
from which we have roughly 
\begin{equation}
\label{a3}
\Gamma(H)\sim\frac{1}{[\Lambda(H)]^{1/(n+1)}},
\end{equation}
and $\Gamma$ is a convex increasing function of $H$ until by our choice of a cut-off it levels out when $\Lambda$ reaches $\Lambda_\infty$.

With the simplifications of \eqref{a2}, \eqref{a1} takes the form of a nonlinear diffusion equation
\begin{equation}
\label{a4}
\left(H^3\right)_x=\nu\frac{\partial}{\partial y}\left[\frac{\Gamma'(H)H^3}{\Gamma(H)^2}\frac{\partial H}{\partial y}\right]+\Gamma(H),
\end{equation} 
which additionally carries the constraint of requiring a prescribed ice flux, which is roughly (with unit ice width and $u\sim\Gamma$)
\begin{equation}
\label{a5}
\int_0^1\Gamma(H)\,dy=1,
\end{equation}
consistent with the assumption of no flux boundary conditions $H_y=0$ on the boundaries $y=0,1$.

Equations of this type with an added integral constraint have been studied by \cite{Fowler:2007aa} and \cite{KyrkeSmith:2014aa}, where it is shown that a stable pseudo-sinusoidal cross-sectional profile emerges at large $x$, a uniform state being unstable. Evidently the wavelength of the resulting pattern is of $O(\sqrt{\nu})$, and the profile is consistent with that of $u_b$ in figure \ref{figA1}. One might query in this figure the odd flattening of the effective pressure profile in the bottom right panel, but we associate this with the somewhat artificial flattening of the $\Lambda(H)$ profile, which effectively acts as a trunctation to an otherwise smooth profile.

This cursory discussion suggests an explanation for the apparent equality of stream width and spacing, and also suggests that these should be $\propto\sqrt{\nu}$ as $\nu\to 0$. This latter suggestion is not completely inconsistent with the results of Figure 4, although it is not particularly supported by them; other than vagaries of defining stream width, whether the computational results are in fact steady, and so on, we have no further suggestion to offer on this (slight) inconsistency.


%
%
%
%
%
%
%

\section*{Acknowledgements}
We thank Stephen Price, Martin Truffer and one anonymous reviewer for very helpful and constructive comments on the manuscript. This work was supported by the Natural Environment Research Council [NE/I528485/1] while Kyrke-Smith was at the University of Oxford. Katz is grateful to the Leverhulme Trust for support. Fowler\ acknowledges the support of the Mathematics Applications Consortium for Science and Industry ({\tt www.macsi.ul.ie}) funded by the Science Foundation Ireland grant 12/1A/1683. Numerical simulations were performed on clusters at the Advanced Research Computing facility of the University of Oxford. Source code for results shown in this paper is available by contacting the corresponding author directly.

\end{document}